\documentclass[10pt,onecolumn,a4paper]{IEEEtran}
\usepackage{graphicx}
\renewcommand{\baselinestretch}{1.6}
\newcommand{\figura}[3]
{
\begin{figure}
  \centering
 \includegraphics[width=14cm]{#1}
  \caption{#2}\label{#3}
\end{figure}
}

\title{LDPC-Based Iterative Algorithm for Compression of Correlated Sources at Rates
Approaching the Slepian-Wolf Bound}
\author{Fred~Daneshgaran,
Massimiliano~Laddomada,
and Marina~Mondin,
\thanks{
This work has been supported by Euroconcepts S.r.l.
(http://www.euroconcepts.it)}%
\thanks{F. Daneshgaran is with the ECE Dept.,
Calif. State Univ., Los Angeles, USA.}
\thanks{M. Laddomada and M. Mondin
are with the Dipartimento di Elettronica, Politecnico di Torino,
Italy.} }
\begin{document}
\maketitle
\begin{abstract}
This article proposes a novel iterative algorithm based on Low
Density Parity Check (LDPC) codes for compression of correlated
sources at rates approaching the Slepian-Wolf bound. The setup
considered in the article looks at the problem of compressing one
source at a rate determined based on the knowledge of the mean source correlation
at the encoder, and employing the
other correlated source as side information at the decoder which
decompresses the first source
based on the estimates of the actual correlation.
We demonstrate that depending on
the extent of the actual source correlation estimated through an
iterative paradigm, significant compression can be obtained
relative to the case the decoder does not use the implicit
knowledge of the existence of correlation.

\end{abstract}
%
%
\section{Introduction}
Consider two independent identically distributed (i.i.d.) discrete binary memoryless
sequences of length $k$, $X=[x_{1},x_{2},\ldots ,
x_{k}]$ and $Y=[y_{1},y_{2},\ldots , y_{k}]$, where pairs of components
$(x_i ,y_i)$ have joint probability mass function $p(x,y)$. Assume
that the two sequences are generated by two transmitters which do not
communicate with each other, and that both sequences have to be
jointly decoded at a common receiver. Slepian and Wolf \cite{slepian}
demonstrated that the achievable rate region for this problem
(i.e., for perfect recovery of both sequences at a joint
decoder), is the one identified by the following set of equations
imposing constraints on the rates $R_X$ and $R_Y$ at which
both correlated sequences are transmitted:
\begin{equation}\label{slep_wolf_achievable_region}
\left \{
\begin{array}{l}
R_X\ge H(X|Y),\\
R_Y\ge H(Y|X),\\
R_X+R_Y\ge H(X,Y)
\end{array} \right.
\end{equation}
whereby $H(X|Y)$ is the conditional entropy of source $X$ given
source $Y$, $H(Y|X)$ is the conditional entropy of source $Y$
given source $X$, and $H(X,Y)$ is the joint entropy. A pictorial
representation of this achievable region is given in
Fig.~\ref{slep_wolf}-a. 

In this article, we focus on trying to achieve the corner points $A$
and $B$ in Fig.~\ref{slep_wolf}-a, since any other point between
these can be achieved with a time-sharing approach \cite{slepian}.
In particular, we focus on the architecture shown in
Fig.~\ref{slep_wolf}-b in which we assume that one of the two
sequences, namely $X$ in our framework, is independently encoded
with a source encoder that has the knowledge of the mean correlation between
the sources $X$ and $Y$. We assume that sequence $Y$ is compressed
up to its source entropy $H(Y)$ and is known at the joint decoder
as side information, and our aim is at compressing sequence $X$
with a rate $R_X$ as close as possible to its conditional entropy
$R_X\ge H(X|Y)$ in order to achieve the corner point $A$ in
Fig.~\ref{slep_wolf}-a. The decoder tries to decompress the
sequence $X$, in order to obtain an estimate $\widehat{X}$, by
employing $Y$ as side information. As shall be seen shortly, the decoder
has an implicit knowledge of mean correlation between sources from the block
length of the encoded sequence. It
estimates the actual correlation between the two sequences through an iterative algorithm which
improves the decoding reliability of $X$.
Obviously, our solution to joint source coding at point $A$ is directly
applicable to point $B$ by symmetry. The overall rate of transmission of both sequences
is greater than $H(Y)+H(X|Y)=H(X,Y)$.

With this background, let us provide a quick survey of the recent
literature related to the problem addressed in this article. This
survey is by no means exhaustive and is meant to simply provide a
sampling of the literature in this area.

In \cite{aaron}, the authors show that turbo codes can allow one to
come close to the Slepian-Wolf bound in lossless distributed
source coding. In \cite{bajcsy,bajcsy3}, the authors propose a
practical coding scheme for separate encoding of the correlated
sources for the Slepian-Wolf problem. In \cite{garcia}, the
authors propose the use of punctured turbo codes for compression
of correlated binary sources whereby compression has been achieved
via puncturing. The proposed source decoder utilizes an iterative
scheme to estimate the correlation between two different sources.
In \cite{garcia_2}, punctured turbo codes have been applied to the
compression of non-binary sources.

Paper \cite{liveris5} deals with the use of parallel and serial
concatenated convolutional codes as source-channel codes for the
transmission of a memoryless binary sequence with side information
at the decoder, while in \cite{liveris3,liveris_c} the authors
propose a practical coding scheme based on LDPC codes for separate
encoding of the correlated sources for the Slepian-Wolf problem.
The problem of Slepian-Wolf correlated source coding over noisy
channels has been dealt with in papers
\cite{garcia_turbo}-\cite{daneshgaran_tcom}.

Relative to the cited articles, the main novelty of the present work
may be summarized as follows: 1) in [5] and [9] the encoder and decoder must both know
the correlation between the two sources. We assume knowledge of mean correlation
at the encoder. The decoder has implicit knowledge of this via observation of
the length of the encoded message. It iteratively estimates the {\em actual} correlation
observed and uses it during decoding; 2) our algorithm can be used with any pair
of systematic encoder/decoder without modifying the encoding and decoding algorithm;
3) the proposed algorithm is very efficient in terms of the required number
of LDPC decoding iterations. We use quantized integer LLR values (LLRQ)
and the loss of our algorithm for using integer LLRQ metrics is quite
negligible in light of the fact that it is able to guarantee performance better
than that reported in [5] and [9]
(where, to the best of our knowledge, authors use floating point metrics)
as exemplified by the results shown in table II below; 4) we utilize post
detection correlation estimates to generate extrinsic information,
which can be applied to any already employed decoder without any modification;
and 5) we do not use any interleaver between the sources at the transmitter.
Using the approach of [5] in a network, information about interleavers used
by different nodes must be communicated and managed. This is not trivial in a
distributed network such as the internet.
Furthermore, there is a penalty in terms of delay that is incurred.
%
\section{Architecture of the LDPC-based Source Encoder}
This section focuses on the source encoder used for source
compression. LDPC coding is essential to achieving performance
close to the theoretical limit in \cite{slepian}.
The LDPC matrix \cite{eff_encod_urbanke} for encoding each source
is considered as a systematic $(n,k)$ code. The codes used need to
be systematic for the
decoder to exploit the estimated correlation between $X$ and $Y$
directly. Each codeword $C$ is composed of a systematic part $X$, and a
parity part $Z$ which together form $C=[X,Z]$. With this setup and
given the parity check matrix $H^{n-k,n}$ of the LDPC code, it is
possible to decompose $H^{n-k,n}$ as follows:
\begin{equation}\label{H_matrix}
H^{n-k,n}=(H^{X},H^{Z})
\end{equation}
whereby $H^{X}$ is a $(n-k)\times (k)$ matrix specifying the
source bits participating in check equations, and $H^{Z}$ is a
$(n-k)\times(n-k)$ matrix of the form:
\begin{equation}
H^{Z}=\left(
\begin{array}{ccccc}
1 & 0 & \ldots  & 0 & 0     \\
1 & 1 & 0       & \ldots & 0 \\
0 & 1 & 1       & 0 & \ldots \\
\ldots &\ldots & \ldots &\ldots&\ldots \\
0& \ldots &0&1&1
\end{array} \right).
\label{h_echelon}
\end{equation}
The choice of this structure for $H$, also called staircase LDPC
(for the double diagonal of ones in $H^{Z}$), has been motivated by the fact
that aside from being systematic, we obtain a LDPC code which is
encodable in linear time in the codeword length $n$. In
particular, with this structure, the encoding operation is as
follows:
\begin{equation}
z_i=\left \{
\begin{array}{ll}
\left[\sum_{j=1}^{k} x_j\cdot H^X_{i,j}\right]~(mod ~2), & i=1 \\
\left[z_{i-1}+\sum_{j=1}^{k} x_j\cdot H^X_{i,j}\right]~(mod~2),
&i=2,.,n-k
\end{array} \right.
\label{h_parities}
\end{equation}
where $H^X_{i,j}$ represents the element $(i,j)$ of the matrix
$H^X$, and $x_j$ is the $j$-th bit of the source sequence $X$.

Source compression is performed as follows; considering the scheme
shown in Fig.~\ref{slep_wolf}-b, we encode the length $k$ source
sequence $X$ and transmit on a perfect channel only the parity
sequence $Z$, whose bits are evaluated as in (\ref{h_parities}).
The rate guaranteed by such an encoder is $R_X=\frac{n-k}{k}$. In
relation to the setup shown in Fig.~\ref{slep_wolf}-b, the
Slepian-Wolf problem reduces to that of encoding the source $X$
with a rate $R_X$ as close to $H(X|Y)$ as possible (i.e., $R_X\ge
H(X|Y)$). The objective of the joint decoder is to recover
sequence $X$ by employing the correlated source $Y$ (considered as
side information at the decoder), and the estimates of the actual
correlation between the sources $X$ and $Y$ obtained in an
iterative fashion.

We consider the following model in order to follow the same
framework pursued in the literature \cite{garcia,liveris3}:
\begin{equation}\label{correlat_model}
P(x_{j}\ne y_{j})=p, ~\forall j=1,\ldots,k
\end{equation}
In light of the considered correlation model, and noting that the
sequence $Y$ is available losslessly at the joint decoder
($R_Y=1$), the theoretical limit for lossless compression of $X$
is $R_X\ge H(X|Y)=H(p)$, whereby $H(p)$ is the binary entropy
function.

Note that the encoder needs to know the mean correlation so as to choose a
rate close to $H(p)$. It does so, by keeping $k$ constant while choosing
$n$ appropriately. We use the term mean correlation, because in any actual
setting, the exact correlation between the sequences may be varying about
the mean value. Hence, it is beneficial if the decoder estimates the actual
correlation value from observations itself. While no side information about the
rate is communicated to the decoder, the decoder knows the mean correlation implicitly
from the knowledge of block length $n$.
\section{Joint Iterative LDPC-Decoding of Correlated Sources}
The architecture of the iterative joint decoder for the
Slepian-Wolf problem is depicted in Fig.~\ref{slep_wolf}-c. Its
goal is to determine the best estimate $\widehat{X}$ of the source
$k$-sequence $X$, by starting from the received parity bit
sequence $Z$ of length $(n-k)$.

Based on the notation above, we can now develop the algorithm for
exploiting the source correlation in the LDPC decoder. Consider a
$(n,k)$-LDPC identified by the matrix $H^{(n-k,n)}$ as expressed
in (\ref{H_matrix}). Note that we only make reference to maximum
rank matrix $H$ since the particular structure assumed for $H$
ensures this. In particular, the double diagonal on the parity
side of the $H$ matrix always guarantees that the rank of $H$ is
equal to the number of its rows, i.e., $n-k$.

For conciseness, we will present only the modifications
to the classical belief-propagation algorithm. The main modification
concerns the initialization step whereby in our setup, each
bit-node is assigned an a-posteriori LLR as follows:
\begin{equation}\label{llr_channel_s}
L\left(u_j\right)=\left \{
\begin{array}{ll}
\log\left(\frac{P(x_j=1|y_j)}{P(x_j=0|y_j)}\right)=\left(2y_j-1\right)\alpha^{(i)}, & j=1,\ldots,k \\
\left(2z_j-1\right),                & j=k+1,\ldots,n
\end{array} \right.
\end{equation}
where $\alpha^{(i)}=\log\left(\frac{p^{(i)}}{1-p^{(i)}}\right)$ is
the correction factor taking into account the estimated
correlation between sequences $X$ and $Y$ at global iteration $i$.
Note that this term derives from the correlation model adopted in
this paper as expressed in (\ref{correlat_model}), in which the
correlation between any bit in the same position in the two
sequences $X$ and $Y$ is seen as having been produced by an
equivalent binary symmetric channel with transition probability
$p$.

The architecture of the iterative joint decoder is depicted in
Fig.~\ref{slep_wolf}-c. We note that there are two stages of
iterative decoding. Index $i$ denotes a {\em global iteration}
whereby during each global iteration, the updated estimate of the
actual source correlation obtained during the previous global iteration
is passed on to the belief-propagation decoder that performs {\em
local iterations} with a pre-defined stopping criterion and/or a
maximum number of local decoding iterations.

Let us elaborate on the signal processing involved. In particular,
as before let $x$ and $y$ be two correlated binary random
variables which can take on the values $\{0,1\}$ and let
$r=x\oplus y$. Let us assume that random variable $r$ takes on the
values $\{0,1\}$ with probabilities $P(r=1)=p_r$ and
$P(r=0)=1-p_r$.

The correction factor $\alpha^{(i)}$ at global iteration $(i)$ is
evaluated as follows,
\begin{equation}\label{llrz}
\alpha^{(i)}=\log\left(\frac{p_{\hat{r}}}{1-p_{\hat{r}}}\right) ,
\end{equation}
by counting the number of places in which $\widehat{X}^{(i)}$ and
$Y$ differ, or equivalently by evaluating the Hamming weight
$w_H(.)$ of the sequence
$\widehat{R}^{(i)}=\widehat{X}^{(i)}\oplus Y$ whereby, in the
previous equation, $p_{\hat{r}}=\frac{w_H(\widehat{R}^{(i)})}{k}$.
In the latter case, by assuming that the sequence
$\widehat{R}=\widehat{X} \oplus Y$ is i.i.d., we have:
\begin{equation}\label{llrz2}
\alpha^{(i)}=\log\left(\frac{w_H(\widehat{R}^{(i)})}{k-w_H(\widehat{R}^{(i)})}\right)
\end{equation}
where $k$ is the source block size. Above, letters highlighted
with $\widehat{.}$ are used to mean that the respective parameters
have been estimated.

Formally, the iterative decoding algorithm can be stated as
follows:
%
%
%
\begin{enumerate}
\item Set the log-likelihood ratios $\alpha^{(0)}$ to proper initial values based
on the knowledge of the mean source correlation (see
Fig.~\ref{slep_wolf}-c). Compute the log-likelihood ratios for any
bit node using~(\ref{llr_channel_s}).

\item For each global iteration $i=1,\ldots,M$, do the following:
\begin{enumerate}
    \item perform belief-propagation decoding on the parity bit sequence $Z$
    by using a predefined maximum number of local iterations, and the side information
    represented by the correlated sequence $Y$ along with the correction factor $\alpha^{(i-1)}$;
    \item Evaluate $\alpha^{(i)}$ using~(\ref{llrz2});
    \item If $\left|\alpha^{(i)}-\alpha^{(i-1)}\right|\ge 10^{-4}$ go back to (a) and continue iterating, else exit.
\end{enumerate}
\end{enumerate}

Step c) in the previous code fragment is used in order to speed-up
the overall iterative algorithm. Extensive tests we conducted
suggested that the threshold value of $10^{-4}$ may be used for
this purpose. Obviously, one can keep iterating until the last
global iteration as well.
\subsection{Overview of Integer-Metrics Belief-Propagation Decoder}
In this section, we briefly describe the LDPC decoder working with
integer LLRs. This approach leads to efficient belief-propagation
decoding. We begin by quantizing any real LLR (denoted LLRQ after
quantization) employed in the initialization phase of the
belief-propagation decoder in (\ref{llr_channel_s}), using the
following transformation:
\begin{equation}\label{metrics_quantiz}
LLRQ=\left \{
        \begin{array}{ll}
        \left\lfloor 2^q L(u_j)+0.5 \right\rfloor, & j=1,\ldots,k \\
        \left\lfloor 2z_j-1\right\rfloor \cdot S,  & j=k+1,\ldots,n
        \end{array} \right.
\end{equation}
whereby $\left\lfloor\cdot \right\rfloor$ stands for rounding to
the smaller integer in the unit interval in which the real number
falls, $ L(u_j)$ is the real LLR, $S$ is a suitable scaling
factor, and $q$ is the precision chosen to represent the LLR with
integer metrics. In our belief-propagation decoder, we use $q=3$,
which guarantees a good trade-off between BER performance and
complexity of the decoder implementation. The scaling factor $S$
is the greatest integer metric processed by the iterative decoder.
In our set-up, we use $S=10000$. Note that such a scaling factor
depends on the practical implementations of the belief-propagation
decoder. Suffice it to say that in our setup, $S$ gives high
likelihood to the parity bits $z_j,~\forall j=k+1,\ldots,n$, since
they are transmitted through a perfect channel to the decoder.
\section{Simulation Results and Comparisons}
\label{sim_results}
We have simulated the performance of our proposed iterative joint
source decoder. We follow the same framework as in
\cite{garcia,liveris3,liveris_c}.

In the following, we provide sample simulation results associated
with various $(n,k)$ LDPC codes designed with the technique
proposed in \cite{kiao}. In particular, for a fair comparison with
the results provided in \cite{liveris_c}, we designed various LDCP
codes with source block length $k=16400$. The details and the
parameters of the designed LDPCs are given in Table~\ref{ldpcs}.

Parameters given in Table~\ref{ldpcs} are the source block length
$k$, the codeword length $n$, the rate $R_X$ of the source,
expressed as $\frac{n-k}{k}$ (i.e., inverse of the compression
ratio), the average degree $d_v$ of the bit nodes, and the average
degree $d_c$ of the check nodes of the designed LDPCs. Note that,
the encoding procedure adopted in our approach is different from
the one proposed in \cite{liveris_c} in that we source encode $k$
bits at a time and transmit only $n-k$ bits. In \cite{liveris_c},
the authors proposed a source compression which encodes $n$ source
bits at a time, and transmits $n-k$ syndrome bits.

For local decoding of the LDPC codes, the maximum number of local
iterations has been set to $50$, while the maximum number of
global iterations is $5$, even though the stopping criterion
discussed in the previous section has been adopted.

In order to test the proposed algorithm for varying actual
correlation levels, for any given value of mean correlation $p$,
we generate a uniform random variable having mean value equal to
the mean correlation itself and with a maximum variation of
$\Delta p$ around this mean value. We used the following maximum
variations: $\Delta p=0.5,0.2,0.1$\%, and $\Delta p=0.0$\% which
refers to the case in which the correlation value is not variable,
but fixed.

For each data block, we set the actual correlation equal to the mean
correlation plus this perturbation. The decoder iterates to
estimate the actual correlation value which varies around its mean
value from one block to the next. In effect, the parameter $p$ is
iteratively estimated as discussed in the previous section. A
similar approach has been pursued in \cite{garcia} for fixed
correlation level, whereby an
iterative approach is used for the estimation of the correlation
between the two correlated sequences, but employing turbo codes.
%
%

Finally, note that we employ integer soft-metrics as explained in
the previous section, while in \cite{garcia,liveris_c}, to the
best our knowledge, the authors employ real metrics. The algorithm
working on integer metrics is very fast and reduces considerably
the complexity burden required by the two-stage iterative
algorithm (i.e., the local-global combination).

Fig.~\ref{sim_various_it} shows the BER performance of the
proposed iterative decoding algorithm for a maximum of $5$ global
iterations and as a function of the joint entropy between sources
$X$ and $Y$, when the stopping criterion for global iterations is
applied. LDPCs used for encoding are the one labelled $L_3$ and
$L_4$ in Table~\ref{ldpcs} which guarantee compression rates of
$R_X=0.237$ and $R_X=0.189$, respectively. LDPC labelled $L_3$ is
used at mean values of $p$ equal to $0.025$, while LDPC $L_4$ is
adopted for a mean correlation of $0.015$. From
Fig.~\ref{sim_various_it} one clearly sees that LDPC decoding does
not converge when the decoder does not iterate for estimating the
actual value of $p$, but uses only its mean value for setting
the extrinsic information. Notice also
that the performances of the iterative decoder when the
correlation value is fixed (curves labelled $\Delta p=0.0$ in
Fig.~\ref{sim_various_it}), are very close to the case in which
the actual correlation value varies within $\Delta p=0.1$\% from
the mean value.

Similar considerations can be deduced from
Fig.~\ref{sim_various_it_complete} which shows the BER performance
of the proposed iterative decoding algorithm when using LDPCs
labelled $L_1$ and $L_2$ in Table~\ref{ldpcs} which guarantee
compression rates of $R_X=0.597$ and $R_X=0.365$, respectively.
LDPC labelled $L_1$ is used at mean correlation equal to $0.1$,
while LDPC $L_2$ is used with a mean correlation of $0.05$. Note that
the performance degrades as $\Delta p$ increases since the encoder
works further away from its optimal operating point.

Finally, we evaluated the average number of global iterations
performed by the iterative algorithm when the stopping criterion
on global iterations is employed during decoding. Simulation
results show that when the LDPC decoder works at BER levels below
$10^{-5}$, the average number of global iterations equals $1.2$,
thus guaranteeing a very efficient iterative approach to the
co-decompression problem. In other words, an overall average
number of $80$ LDPC decoding iterations suffices to obtain good
BER performance.

The results on the compression achieved with the proposed
algorithm are shown in Table~\ref{number_of_iterations} for the
case in which the correlation value is fixed. The first row shows
the fixed correlation parameter assumed, namely,
$p=P(x_j\ne y_j),~\forall j=1,\ldots,k$ in our model.
The second row shows the joint entropy limit for various values of
the fixed correlation parameter $p$. The third and fourth rows show
the results on source compression presented in papers
\cite{garcia,liveris_c}, while the last row presents the results
on compression achieved with the proposed algorithm employing a
maximum of 5 global iterations in conjunction with using the
stopping criterion noted in the previous section. As in
\cite{liveris_c}, we assume error free compression for a target
Bit Error Rate (BER) $10^{-6}$. Note that statistic of the results
shown has been obtained by counting 30 erroneous frames.

From Table~\ref{number_of_iterations} it is evident that
significant compression gains with respect to the theoretical
limits can be achieved as the correlation between sequences
$X$ and $Y$ increases.
%
%
%
%
%
%
%
%
%
%
%
%
%
%
%
\section{Conclusions}
In this article we have presented a novel iterative joint decoding
algorithm based on LDPC codes for the Slepian-Wolf problem of
compression of correlated information sources.
In the considered scenario, two correlated sources communicate with
a common receiver. The first source is compressed by transmitting
the parity check bits of a systematic LDPC encoded codeword. The
correlated information of the second source is employed as side
information at the receiver and used for decompressing and
decoding of the first source.
The crucial observation is that LDPC decoding does
not converge when the decoder does not iterate for estimating the
actual value of $p$, but uses instead its mean value which is
assumed to be implicitly known. Both the
iterative decoding algorithm and the cross-correlation estimation
procedure have been described in detail.
Simulation results suggest that relatively large compression gains
are achievable at relatively small number of global iterations
specially when the sources are highly correlated.
\clearpage
\figura{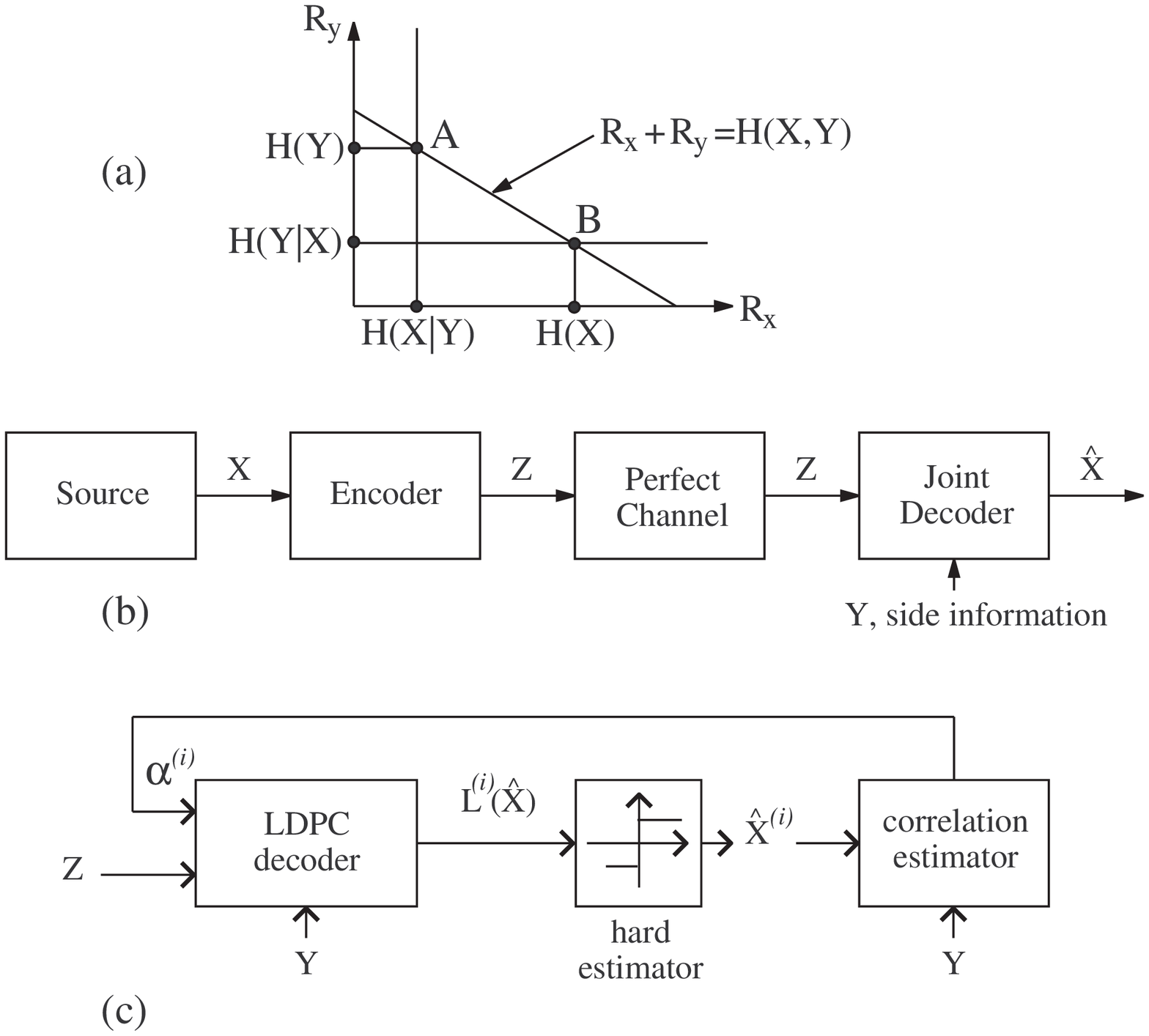}{Rate region for Slepian-Wolf encoding (a).
Architecture of the encoder and joint decoder for the Slepian-Wolf
problem (b). Architecture of the Iterative Joint decoder of
correlated sources (c).}{slep_wolf}
\clearpage
\figura{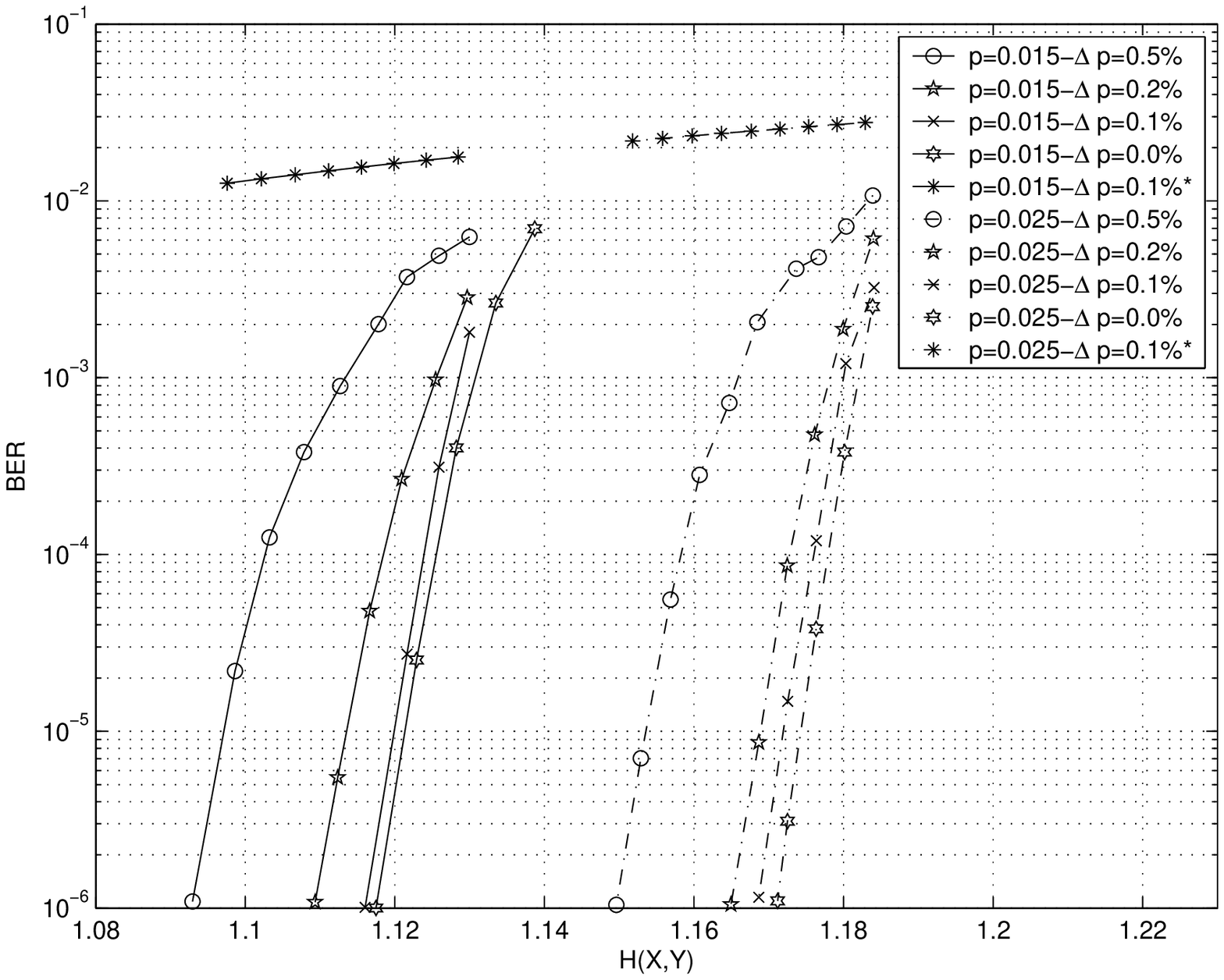}{BER performance of the proposed iterative
decoding algorithm for a maximum of $5$ global iterations as a
function of the joint entropy between sources $X$ and $Y$, when
the stopping criterion for global iterations is applied. Results
refer to the LDPCs labelled $L_4$ and $L_3$ in Table~\ref{ldpcs}.
The legend shows the mean correlation value $p$ and the maximum
value of the correlation variation with respect to the mean value.
Curves labelled with $*$ refer to the ones obtained without the
iterative paradigm, using the mean correlation
value.}{sim_various_it}
\clearpage
\figura{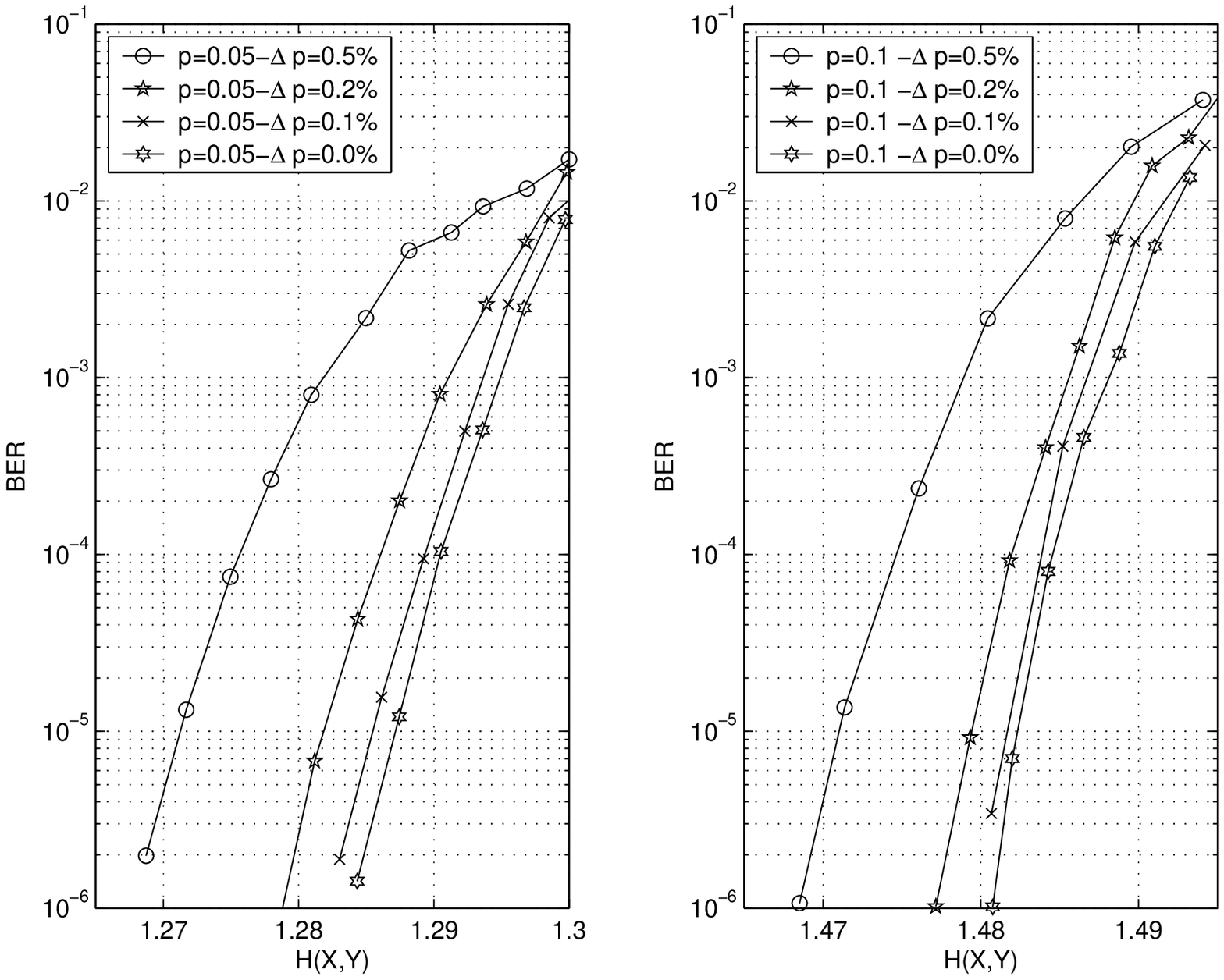}{BER performance of the proposed iterative
decoding algorithm for a maximum of $5$ global iterations as a
function of the joint entropy between sources $X$ and $Y$, when
the stopping criterion for global iterations is applied. Results
refer to the LDPCs labelled $L_2$ (left subplot) and $L_1$ (right
subplot) in Table~\ref{ldpcs}. The legend shows the mean
correlation value $p$ and the maximum value of the correlation
variation with respect to the mean
value.}{sim_various_it_complete}
\clearpage
\begin{table}\large
\caption{Parameters of the designed LDPCs.}
\begin{center}
\begin{tabular}{l|c|c|c|c|c} \hline \hline
LDPC & $k$ & $n$ & $R_X$ & $d_v$ & $d_c$
\\\hline \hline

%

$L_1$ & 16400 & 26200 & 0.597 & 3  & 8\\\hline

$L_2$ & 16400 & 22400 & 0.365 & 3.21  & 12

%

\\\hline

$L_3$ & 16400 & 20300 & 0.237 & 3.45  & 18

%
%
%

\\\hline

$L_4$ & 16400 & 19500 & 0.189 & 3.0  & 19

\\\hline\hline

\end{tabular}
\label{ldpcs}
\end{center}
\end{table}
\begin{table}\large
\caption{Compression rate performance of the iterative algorithm
for various joint entropies.}
\begin{center}
\begin{tabular}{l|c|c|c|c} \hline\hline

$p$  &   0.015   & 0.025     &  0.05      & 0.1 \\\hline\hline

$H(p)+1$&   1.112 &   1.169   &  1.286     & 1.469
\\\hline

$R$~\cite{garcia}&  -&   1.31 &  1.435     & 1.63
\\\hline

$R$~\cite{liveris_c}&  -& 1.276  &  1.402     & 1.60
\\\hline

$R=R_X+R_Y$ &  1.189  -$L_4$      &    1.237 -$L_3$  & 1.365-$L_2$
& 1.597-$L_1$

\\\hline \hline
\end{tabular}
\label{number_of_iterations}
\end{center}
\end{table}
\end{document}